\documentclass[journal,10pt,twocolumn]{IEEEtran}

\newtheorem{proposition}{Proposition}
\IEEEoverridecommandlockouts
\usepackage{amsmath,amssymb,amsfonts,float,graphicx,steinmetz,bm,subfigure,multirow,url,color,xcolor,etoolbox,ifthen,cite}
\usepackage{mathtools}
\usepackage{textcomp}

\DeclarePairedDelimiter\floor{\lfloor}{\rfloor}
\DeclareMathSizes{10}{9.5}{6}{5}

\def\undb#1{\mbox{\bf{#1}}}

\begin{document}  

\title{ \huge
Optimal Grouping Strategy for Reconfigurable Intelligent Surface Assisted Wireless Communications
\thanks{
\textbf{This work has been submitted to the IEEE for possible publication. Copyright may be transferred without notice, after which this version may no longer be accessible.} This work was supported by the Hong Kong Research Grants Council (grant
number C6012-20G). {\em Corresponding Author: Shanpu Shen.}
}
}

\author{
Neel Kanth Kundu, {\em Student Member, IEEE}, Zan Li, {\em Student Member, IEEE}, Junhui Rao, {\em Student Member, IEEE}, Shanpu Shen, {\em Member, IEEE},
Matthew R. McKay, {\em Fellow, IEEE}, and Ross Murch, {\em Fellow, IEEE}
\thanks{
The authors are with the Department of
Electronic and Computer Engineering (ECE), The Hong Kong University of
Science and Technology (HKUST), Clear Water Bay, Kowloon, Hong Kong.  R. Murch is also with the Institute for Advanced Study, HKUST.

M. R. McKay was with the Department of ECE, HKUST. He is now with the Department of Electrical and Electronic Engineering, University of Melbourne, Melbourne, Australia.
(e-mails:\{nkkundu, zligq, jraoaa, sshenaa\}@connect.ust.hk,  m.mckay@ust.hk,  eermurch@ust.hk).}
\vspace{-0.2in}
}
\markboth{This work has been submitted to the IEEE for possible publication. Copyright may be transferred without notice.}{This work has been submitted to the IEEE for possible publication. Copyright may be transferred without notice.}
\maketitle
\begin{abstract}

The channel estimation overhead of reconfigurable intelligent surface (RIS) assisted communication systems can be prohibitive. Prior works have demonstrated via simulations that grouping neighbouring RIS elements can help to reduce the pilot overhead and improve achievable rate. In this paper, we present an analytical study of RIS element grouping. We derive a tight closed-form upper bound for the achievable rate and then maximize it with respect to the group size. Our analysis reveals that more coarse-grained grouping is important-when the channel coherence time is low (high mobility scenarios) or the transmit power is large. We also demonstrate that optimal grouping can yield significant performance improvements over simple `On-Off' RIS element switching schemes that have been recently considered.



\end{abstract}


\IEEEpeerreviewmaketitle
\vspace{-0.1in}
\section{Introduction}
Reconfigurable intelligent surfaces (RIS) present a new antenna technology for 6G wireless systems \cite{pan2021reconfigurable,basar2019wireless,renzoRISsurvey20,huang2020holographic,shen2020modeling}. RIS is an energy efficient solution that can improve the signal strength at a desired location by intelligently controlling the phase shifts of incoming electromagnetic waves \cite{pan2021reconfigurable,basar2019wireless,renzoRISsurvey20,huang2020holographic,shen2020modeling}. 

Over the past couple of years, researchers have investigated the optimal design of the RIS phase shifts that maximizes some performance metric like achievable rate, signal-to-noise ratio (SNR) or outage probability \cite{huang2019reconfigurable,ruiwu2018intelligent,debbahhuang2018achievable}. These works have shown that RIS can provide significant performance improvement under the assumption of perfect channel state information (CSI). However, in practice, channel estimation for RIS is challenging since it consists of only passive elements without any RF chains and computing capability. Hence, all processing has to be done at the end nodes only. For standard least-squares (LS) based channel estimation schemes, this leads to a linear increase in the pilot overhead with the increasing number of RIS elements, which limits the achievable rate of the system \cite{jensen2019optimal,kundu2021channel,kundu2021large,zheng2019intelligent,zheng2020intelligent}.


Recently, the authors of \cite{zappone2020optimal,kundu2021large} have proposed an `On-Off' based scheme where only an optimized number of RIS elements are switched on in order to reduce the pilot overhead. The solution in \cite{zappone2020optimal} required the instantaneous CSI, whereas the authors of \cite{kundu2021large} derived an analytical expression for the optimal number of RIS elements to be switched on, depending only on the statistical CSI.
In related works, a grouping strategy has been proposed by the authors of \cite{zheng2019intelligent,zheng2020intelligent} for reducing the pilot overhead, where the neighbouring RIS elements in a group share a common phase shift. In \cite{zheng2019intelligent,zheng2020intelligent}, it was demonstrated via simulations that there exists an optimal group number that maximizes the achievable rate of the system. 


In this paper, considering the grouping strategy of \cite{zheng2019intelligent,zheng2020intelligent}, we analytically study the optimal group size that maximizes the achievable rate of the system. Specific contributions we make include: 
\begin{enumerate}
  \item We derive an analytical upper bound on the achievable rate,  which is solely dependent on the statistical channel and system parameters. Our simulation results show that the upper bound is tight for practically feasible system parameters.
  \item We find an approximate closed form expression for the optimal group size that maximizes the upper bound. The analytical expression can guide RIS system design based on the knowledge of statistical system parameters.
  \item The analytical expression reveals that the optimal group size increases when the channel coherence time is low (high mobility scenarios) or the transmit power is large.
  \item We evaluate the performance of the grouping strategy via numerical simulations. Our results reveal that the grouping strategy significantly outperforms the `On-Off' strategy of \cite{kundu2021large}, since the grouping strategy reflects more power by utilizing all elements within a group.
\end{enumerate}
\section{System Model}
A single antenna source (S) node is assumed to be communicating with a single-antenna destination (D) node assisted by an RIS having $K$ passive elements. In order to reduce the channel estimation overhead, a software grouping strategy is employed where the nearby $B$ elements are grouped together such that they introduce the same phase shift \cite{zheng2020intelligent}. The effective number of RIS subgroups is given by $K' = \floor*{\frac{K}{B}}$, where $ \floor*{x} $ truncates the fractional part of $x$.
\subsection{Channel Estimation}
We assume a Rayleigh fading environment such that the small-scale fading channels between S-RIS and D-RIS are denoted by $\undb{h} = [ h_1,h_2, \ldots, h_K  ]^T \in {\mathbb C}^{K} $ and $\undb{g} = [ g_1,g_2, \ldots, g_K  ]^T \in{\mathbb C}^{K} $ respectively, 
whose elements are independent and identically distributed (i.i.d) with, $h_i\sim \mathcal{CN}(0,1), g_i \sim \mathcal{CN}(0,1)$. Moreover, the small-scale fading channel of the direct S-D link is denoted by $h_d \in{\mathbb C}$ with $h_d\sim \mathcal{CN}(0,1) $.
The cascaded S-RIS-D channel is represented as $\undb{v} = [v_1, v_2,\ldots,v_K] =  \undb{h} \odot \undb{g}$. During the channel estimation phase the direct and cascaded channels are estimated at the source node by transmitting pilot signals from the destination node. Assuming a time division duplex mode of operation with channel reciprocity, the estimated channels at the source node are then used for passive beamforming during data transmission. 

The signal received at S during the channel estimation phase is \cite{zheng2020intelligent} 
\begin{equation}
    y_m = \sqrt{P_{\rm tr}} (\sqrt{\beta_d} h_d+ \sqrt{\beta_l} \bm{\phi}_m^T \undb{v}') x_m + n_m \;,
    \label{eq1.1}
\end{equation}
where $x_m \in \mathbb{C},\; |x_m|=1$ denotes the transmitted pilot signal from D, and $y_m \in {\mathbb C}$ denotes the pilot signal received at S during the $m$ th pilot transmission. $P_{{\rm tr}}$ is the power of the pilot signal, $n_m  \sim {\mathcal CN} (0, \sigma^2)$ is the additive noise at S, and $\beta_d,\beta_l$ denote the large scale path loss coefficients of the direct and the cascaded link respectively. Further, $\undb{v}' =[v'_1,\ldots,v'_{K'}]$ is the equivalent cascaded channel arising from software grouping such that the $i$ th element of $\undb{v}'$ is given by \cite{zheng2019intelligent,zheng2020intelligent}
\begin{equation}
    v'_i = \sum_{b=1}^{B} v_{b+(i-1)B} \;,
    \label{eq1.2}
\end{equation}
and $\bm{\phi}_m  = [e^{j \theta_{m,1}}, \ldots, e^{j \theta_{m,K'}} ]^{T} \in {\mathbb C}^{K'}$ is the equivalent phase shift vector of the RIS corresponding to the $K'$ subgroups, where $\theta_{m,k} \in [0, 2\pi]$. In order to design the phase shift vector at the RIS for downlink transmission only the equivalent cascaded channel $\undb{v}'\in {\mathbb C}^{K'\times 1} $ and the direct channel $h_d$ need to be estimated. We assume a channel coherence block of length $T_c$, during which the channels $\undb{h}, \undb{g}$ and $h_d$ remain constant. Let $T_p$ denote the pilot duration such that $T_p<T_c$. Then, a LS estimate of $h_d$ and $\undb{v}'$ can be obtained from the linear measurement model of (\ref{eq1.1}), with a DFT-based phase shift matrix for $\bm{\phi}_m$, $m=1,\ldots,T_p$ \cite{jensen2019optimal}. The LS estimate exists when $T_p \geq K'+1$ \cite[eq.~10]{jensen2019optimal}.
Note that for $B>1$, the pilot overhead has been decreased from $K+1$ (without grouping) to $K'+1$ (with grouping) since $K' = \floor*{\frac{K}{B}}$. Although, the number of degrees of freedom and correspondingly the passive beamforming gain reduces due to grouping, the reduction in pilot overhead balances it such that the overall rate improves.
\subsection{Data Transmission}
The signal received at D during data transmission is
\begin{equation}
     y = \sqrt{P} (\sqrt{\beta_d} h_d+ \sqrt{\beta_l} \bm{\phi}^T \undb{v}') x + n \;,
    \label{ed1}
\end{equation}
where $x$ with ${\mathbb E}[|x|^2]=1$ denotes the information symbol, and $P$ is the transmit power during data transmission.
Further, $n \sim \mathcal{CN}(0,\sigma^2) $ is the additive noise at D, and $\bm{\phi} = [\phi_1, \phi_2,\ldots \phi_{K'}]^T \in{\mathbb C}^{K'}$ with $|\phi_i|=1, \;  \forall \; i=1,\ldots,K' $ is the phase shift introduced by the $K'$ subgroups of the RIS. If perfect CSI is available, the received SNR is maximized for \cite[eq.~7]{gao2020unsupervised}
\begin{equation}
    \phi_i = e^{j {\rm arg}\left( \frac{h_d}{v_i'} \right) } \,, \forall \; i=1,2,\ldots K' \;,
    \label{eq1.3} 
\end{equation}
where $j=\sqrt{-1}$. However, in practice the phase shifts are designed using the estimates of $h_d, \undb{v}'$. Let $\hat{h}_d, \hat{\undb{v}}'$ denote the least squares (LS) channel estimates obtained from \cite[eq.~10]{jensen2019optimal}. Then the phase shift vector is computed in practice as
\begin{equation}
    \hat{\phi_i} = e^{j  {\rm arg}\left( \frac{\hat{h}_d}{\hat{v}_i'} \right) }\;, \forall \; i=1,2,\ldots K'.
    \label{eq1.3b}
\end{equation}
\section{Average Achievable Rate}
Here we investigate the system's average achievable rate by incorporating the pilot overhead. Since the pilot overhead for the LS channel estimator is $T_p = K'+1$, the achievable rate is given by \cite{massivemimobook}
\begin{equation}
    R = \left(1-\frac{K'+1}{T_c} \right) {\mathbb E}\left[ \log_2\left(1+ \gamma | \sqrt{\beta_d} h_d + \sqrt{\beta_l} \bm{\hat{\phi}}^T \bm{v'}  |^2 \right) \right]  \;,
    \label{eq1.4}
\end{equation}
where $\gamma = P/\sigma^2$ denotes the transmit SNR. We derive a closed-form upper bound expression for $R$. 
Similar to \cite{kundu2021large}, we use a bounding approach, since it is difficult to exactly compute $R$ due to the complicated distribution of the residual angles (which appear due to imperfect CSI) \cite{badiu2019communication}. Our simulation results show that the derived upper bound is close to the actual $R$ for practically feasible system parameters.
\begin{proposition} \label{proposition1}
An upper bound on the achievable rate $R$ is given by
\begin{align}
    \Bar{R}& = \left(1-\frac{K'+1}{T_c} \right) \log_2\left(1+ \gamma \left( \xi_1 K'^2+ \xi_2 K' +\beta_d \right) \right)\;,
    \label{lemm1.1}
\end{align}
where $\xi_1 =\beta_lz^2, \;\;
  \xi_2= \beta_l\left( B-z^2\right) + \sqrt{\pi \beta_d\beta_l}z $,
and 
\begin{align}
    z = \frac{\sqrt{\pi}\Gamma\left(B+\frac{1}{2}\right)}{2\Gamma\left( B\right)} \;.
  \label{lemm1.2a} 
\end{align}
\label{l1}
\end{proposition}
\begin{IEEEproof}
We apply two successive upper bounding steps. First, we apply Jensen's inequality in (\ref{eq1.4}) to obtain
\begin{equation}
    R \leq \left(1-\frac{K'+1}{T_c} \right)  \log_2\left(1+ \Bar{\gamma} {\mathbb E}\left[| \sqrt{\beta_d} h_d + \sqrt{\beta_l} \bm{\hat{\phi}}^T \undb{v}'  |^2\right] \right) \;.
    \label{lemm1.3} 
\end{equation}
Next, we replace $\hat{\bm{\phi}}$ with $\bm{\phi}$ to further upper bound $R$ \cite{kundu2021large}.
Using (\ref{eq1.3}), the ${\mathbb E}\left[ \cdot\right] $ inside the logarithm of (\ref{lemm1.3}) is given by
\begin{align}
   \beta_d \underbrace{{\mathbb E} \left[ |h_d|^2 \right]}_{z_1}  + \beta_l \underbrace{{\mathbb E} \left[ \left(\sum_{i=1}^{K'} |v'_i| \right)^2 \right]}_{z_2} + 2\sqrt{\beta_d \beta_l} \underbrace{{\mathbb E} \left[ \left( |h_d|\sum_{i=1}^{K'} |v'_i| \right) \right]}_{z_3}.
 \label{e1.6}
\end{align}
It follows trivially that $z_1=1$, whereas $z_2$ is given by
\begin{align}
    z_2 &= \sum_{i=1}^{K'}{\mathbb E}\left[|v'_i|^2 \right] + \sum_{i=1}^{K'}{\mathbb E}\left[ |v'_i| \right] \left( \sum_{\substack{ j=1 \\ j \neq i }}^{K}{\mathbb E}\left[|v'_j| \right] \right) \nonumber \\
    & = K' z_4 + K'(K'-1)z_5^2 \;,
    \label{e1.7}
\end{align}
where
\begin{align}
     z_4={\mathbb E}\left[ |v'_i|^2  \right] \,, \, \textnormal{and} \; z_5  = {\mathbb E}\left[ |v'_{i}| \right]\;.
     \label{e1.8}
\end{align}
Since $v'_i$ are i.i.d. we can evaluate $z_4, z_5$ by choosing $i=1$. Using (\ref{eq1.2}) in (\ref{e1.8}) we obtain 
\begin{align}
  z_4 &=  {\mathbb E}\left[ \left| \sum_{b=1}^{B} v_{b}\right|^2 \right]  = {\mathbb E}\left[ \left( \sum_{b=1}^{B} v_{b} \right) \left( \sum_{b'=1}^{B} v^{*}_{b'} \right) \right] \nonumber \\
  & = {\mathbb E}\left[ \sum_{b=1}^{B} |v_{b}|^2 + \sum_{\substack{ b=1 \\ b \neq b' }}^{B} v_{b} v^{*}_{b'} \right] = B\;,
  \label{e1.10}
\end{align}
where we have used ${\mathbb E}\left[ |v_{b}|^2 \right] = {\mathbb E}\left[ |h_{b}|^2 |g_{b}|^2 \right] = 1 $, ${\mathbb E}\left[ v_{b} \right]={\mathbb E}\left[ h_{b} \right] {\mathbb E}\left[ g_{b} \right] = 0$, and the independence of $v_{b} \;, \forall \; b=1,\ldots,B$. Similarly, $z_5$ can be expressed as
\begin{align}
  z_5 & =  {\mathbb E}\left[ \left| \sum_{b=1}^{B} v_{b} \right| \right] \;.
  \label{e1.11}
\end{align}
We use a conditioning and averaging approach to evaluate $z_5$. Since $v_b=h_bg_b$, for a given $g_b$ the distribution of $v_b$ is given by $v_b|g_b \sim \mathcal{CN}\left( 0, |g_b|^2\right)$. Denote $\delta=\sum_{b=1}^{B} v_b$, then the conditional distribution of $\delta$ is given by
\begin{equation}
    \delta|_{g_b,\ldots g_B} \sim \mathcal{CN}\left( 0, \sum_{b=1}^{B}|g_b|^2\right) \;.
    \label{e1.12}
\end{equation}
For a complex Gaussian random variable $x\sim \mathcal{CN}(0,\rho^2)$, we have $\mathbb{E}\left[|x| \right] = \frac{\rho \sqrt{\pi}}{2} $ \cite{simon2002probability}. Therefore, the conditional mean of $\delta$ is given by
\begin{equation}
    \mathbb{E} \left[ \left| \delta|_{g_1,\ldots g_B} \right| \right] = \frac{\sqrt{\pi}}{2} \sqrt{\sum_{b=1}^{B} |g_b|^2} \;.
    \label{e1.13}
\end{equation}
Using the law of total expectation, $z_5$ can be expressed as
\begin{equation}
    z_5 = \mathbb{E}_{g_1,\ldots g_B} \left[ \frac{\sqrt{\pi}}{2} \sqrt{\sum_{b=1}^{B} |g_b|^2}   \right] \;.
    \label{e1.14}
\end{equation}
Denoting $\alpha = \sum_{b=1}^{B} |g_b|^2  $, and using properties of the Rayleigh distribution, it is easy to verify that $\alpha$ follows a Gamma distribution with shape parameter $B$ and scale parameter $1$. Thus, $\mathbb{E}\left[\sqrt{\alpha}\right]$ can be evaluated as
\begin{align}
 \mathbb{E}\left[\sqrt{\alpha}\right] & = \int_{0}^{\infty} \sqrt{\alpha} f(\alpha) d\alpha  \nonumber \\
 & = \int_{0}^{\infty} \sqrt{\alpha} \frac{\alpha^{B-1} e^{-\alpha} }{\Gamma\left( B\right)} d\alpha =\frac{\Gamma\left(B+\frac{1}{2}\right)}{\Gamma\left( B\right)} \;,
 \label{e1.15}
\end{align}
where the last equality follows from the definition of the Gamma function. Hence, $z_5$ is obtained by using (\ref{e1.15}) in (\ref{e1.14}).
Finally, using  (\ref{e1.10}), (\ref{e1.7}) and $z_3 =0.5\sqrt{\pi}K' z_5$ in (\ref{e1.6}), the upper bound on the achievable rate is given by (\ref{lemm1.1}).
\end{IEEEproof}
\section{Optimal RIS Grouping}
In this section, we find the optimal group size $B$ that maximizes the achievable rate by utilizing the result from Proposition \ref{l1}. We want to solve the following optimization problem
\begin{equation}
     B^{\star} = \underset{1\leq B \leq K}{ {\rm arg \; max}} \;  \Bar{R}(B)
    \label{eq4.1}
\end{equation}
with $\Bar{R}(B)$ given by
\begin{equation}
     \Bar{R}(B) = \left(1-\frac{\frac{K}{B}+1}{T_c} \right) \log_2\left(1+ \gamma \left( \xi_1 \left(\frac{K}{B} \right)^2+ \xi_2 \frac{K}{B} +\beta_d \right) \right) \;.
     \label{eq4.2}
\end{equation}
Since $B$ can take only positive integer values, the optimal $B^{\star}$ in (\ref{eq4.1}) can be found by using a brute force search over the feasible set $B \in \{1,\ldots,K\}$. However, in order to clearly reveal the effect of different system parameters on the optimal group size, we find an approximate solution of (\ref{eq4.1}) under practically feasible assumptions. 

We first find a power function approximation for $z$ given by (\ref{lemm1.2a}) in order to find an analytical solution for $B^{\star}$. Since $z$ is a ratio of two gamma functions with integer values of $B$, by simple inspection we find that $z$ is a sub-linear function\footnote{A function $f(x)$ is sub-linear if $\underset{x\rightarrow \infty}{ \lim }  \frac{f(x)}{x}=0.$} in $B$. Therefore, we fit a power function of the form $\kappa B^{\eta}$ with $0<\eta<1$ for $z$. Moreover, this power function approximation simplifies the analysis for $B^{\star}$. By simple curve fitting we find that $z$ can be approximated as $z \approx 0.8759 \sqrt{B}$. Due to the complicated expression of $\Bar{R}(B)$ in (\ref{eq4.2}), an approximate closed form solution for $B^{\star}$ can only be found when both the transmit SNR and number of RIS elements are large.

\begin{proposition} \label{thm1}
For large transmit SNR $\gamma$ and large number of RIS elements $K$, the optimal group size $B^{\star}$ can be approximated as
\begin{equation}
    B^{\star} = \floor*{ \frac{K}{T_c-1}W\left( \zeta \gamma(T_c-1)K \right) + \frac{1}{2} } \;,
    \label{eq4.3}
\end{equation}
where $\zeta=2.08\beta_l $, and $W(x)$ is the Lambert's W-function.
\end{proposition}
\begin{IEEEproof}
As $\gamma \to \infty$, $\log_2 (1 + \psi \gamma ) = \log_2 ( \psi \gamma ) + o(1)$, for some constant $\psi$. Hence, using $z\approx 0.8759 \sqrt{B}$ in (\ref{eq4.2}), we obtain
\begin{equation}
    \Bar{R}_1(B) = \left( 1-\frac{1+\frac{K}{B}}{T_c}\right) \log_2\left(1+c\gamma \left(\frac{K^2}{B} \right) \right) + o(1) \;, 
    \label{eq4.4}
\end{equation}
where $c=0.7671\beta_l$. Setting $ \Bar{R}_1'(B)=0$, we obtain
\begin{equation}
    \tau - K\ln{B} = B(T_c-1) \;,
    \label{eq4.5}
\end{equation}
where $\tau=K\ln{e\gamma cK^2}$, and $e$ denotes Euler's number. Denoting $t=\frac{(T_c-1)B}{K}$, and after some simple algebra, (\ref{eq4.5}) can be simplified as
\begin{equation}
    t e^t=\frac{T_c-1}{K}e^{\frac{\tau}{K}} = ce\gamma(T_c-1)K \;.
    \label{eq4.6}
\end{equation}
The solution for $t$ in (\ref{eq4.6}) can be obtained by using Lambert's W-function as \cite{corless1996lambertw}
\begin{equation}
    t = W\left( ce\gamma(T_c-1)K\right) \;.
    \label{eq4.7}
\end{equation}
After some simple algebra and under the constraint of $B$ taking only integer values, $B^{\star}$ is given by (\ref{eq4.3}).
\end{IEEEproof}

The result of Proposition \ref{thm1} can be used to understand the effect of important system parameters on the optimal group size $B^{\star}$. The two important system parameters which affect the group size are the data transmit power $P$ and channel coherence time $T_c$. First, we study the effect of $P$. We note that $W(x)$ is a monotonically increasing function of $x$ for $x>0$ \cite{corless1996lambertw}. Since $\zeta, \gamma ,K >0 $ and $T_c>1$, the argument of $W(\cdot)$ in (\ref{eq4.3}) is positive, therefore as $P$ (and correspondingly $\gamma$) increases, $B^{\star}$ also increases. This phenomenon can also be intuitively understood from the objective function in (\ref{eq4.2}). It can be observed that at higher $\gamma$, the second term in (\ref{eq4.2}) increases logarithmically with decreasing $B$, whereas the effect of higher $B$ is more pronounced in the pre-log factor. Hence, it is beneficial to reduce the pilot overhead by having a smaller $K'$ (equivalently a higher $B^{\star}$).


Further, to understand the effect of the coherence time $T_c$ on $B^{\star}$, we use the identity $\frac{W(x)}{x} = e^{-W(x)}$. Using this, (\ref{eq4.3}) can be equivalently written as
\begin{equation}
    B^{*} = \left \lfloor \zeta \gamma K^2 e^{-W[\zeta \gamma K(T_c-1)]} + \frac{1}{2} \right \rfloor \; .
    \label{eq4.9}
\end{equation}
As before, the argument of $W(\cdot)$ in (\ref{eq4.9}) is positive and increases with $T_c$. However, $e^{-x}$ is a monotonically decreasing function, hence $B^{\star}$ decreases with increasing $T_c$. This can also be intuitively understood from the objective function in (\ref{eq4.2}), since for large $T_c$, the effect of the pre-log factor is reduced and it is desirable to increase the beamforming gain inside the log term by increasing the number of subgroups $K'$ (which corresponds to a smaller group size $B^{\star}$). 

\section{Simulation Results}
We consider a simulation scenario similar to \cite{ruiwu2018intelligent}, where the source node S and the RIS are located on a horizontal line with the distance between them being $d_0= 51$ m. The destination node D is located near the RIS at a vertical distance of $d_v = 2$ m, and the horizontal distance between S and D is $d= 48$ m. First, we check the accuracy of the upper bound of the achievable rate derived in (\ref{lemm1.1}).
\begin{figure}[ht] 
\centering
\subfigure[ ]{%
\includegraphics[width=0.4\textwidth]{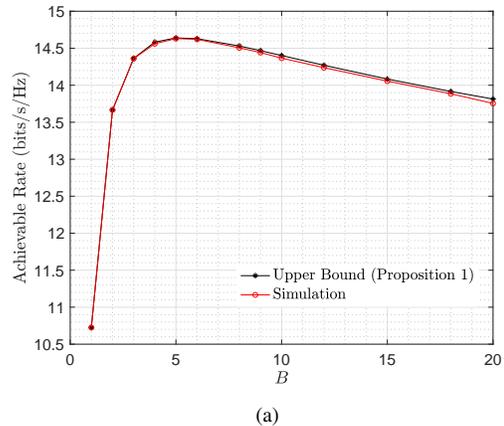}%
\label{fig3.1:a}%
}
\subfigure[ ]{%
\includegraphics[width=0.4\textwidth]{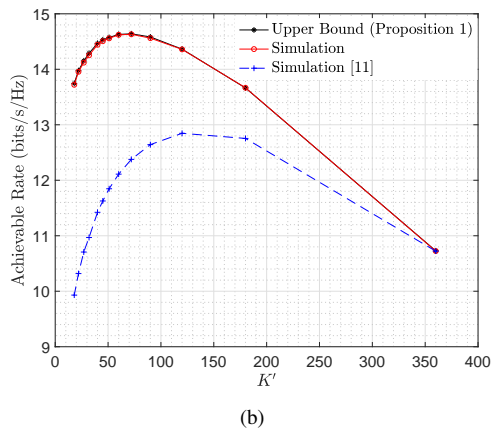}%
\label{fig3.1:b}%
}
\caption{ The plots show the achievable rate $R$ versus (a) number of elements in each subgroup $B$ and (b) number of subgroups $K'$. {\emph{System Parameters:}} $P_{\rm tr}=P= 0$ dB, $K=360$, and $T_c= 900$. We use a path loss exponent of $3.5$ and $2$ for the direct and cascaded channel, respectively. We assume that the path loss is $-30$ dB at a distance of $1$ m, and the receiver noise variance is $\sigma^{2}= -80$ dBm \cite{ruiwu2018intelligent}.
}
\label{fig3.1}
\end{figure}
Fig. \ref{fig3.1} shows the plot of the achievable rate as a function of (a) group size $B$, and (b) number of subgroups $K'$. The plots compare the upper bound from (\ref{lemm1.1}) and the actual achievable rate obtained from Monte-Carlo simulation of (\ref{eq1.4}) with LS channel estimation. We observe that the upper bound is accurate, and the approximate location of $B^{\star}$ (that maximizes the achievable rate) is the same for the simulated curve and the closed-form upper bound.
Thus, the upper bound can be used to find the optimal group size $B^{\star}$. From Fig. \ref{fig3.1:a} we observe that grouping of elements with $B>1$ leads to a better performance as compared to no grouping ($B=1$), since the channel estimation overhead $T_p = \floor{\frac{K}{B}} + 1$ is reduced for $B>1$. For comparison, we also show the achievable rate of the `On-Off' scheme of \cite{kundu2021large} in Fig. \ref{fig3.1:b} where only $K'$ elements are turned on and the rest of the elements have reflection coefficient equal to zero. Note that the grouping scheme of this paper with $K'$ subgroups, and the `On-Off' scheme of \cite{kundu2021large} with only $K'$ elements switched on have the same pilot overhead, however the performance of the grouping method is better than the `On-Off' scheme. The reason is that for the grouping scheme all the elements in a group reflect the incoming signal (albeit with the same phase shift), whereas for the `On-Off' scheme only one element of that group reflects the incoming signal. Therefore, the grouping scheme reflects more power which leads to its enhanced performance.

\begin{figure}[ht] 
\centering
\subfigure[ ]{%
\includegraphics[width=0.5\linewidth]{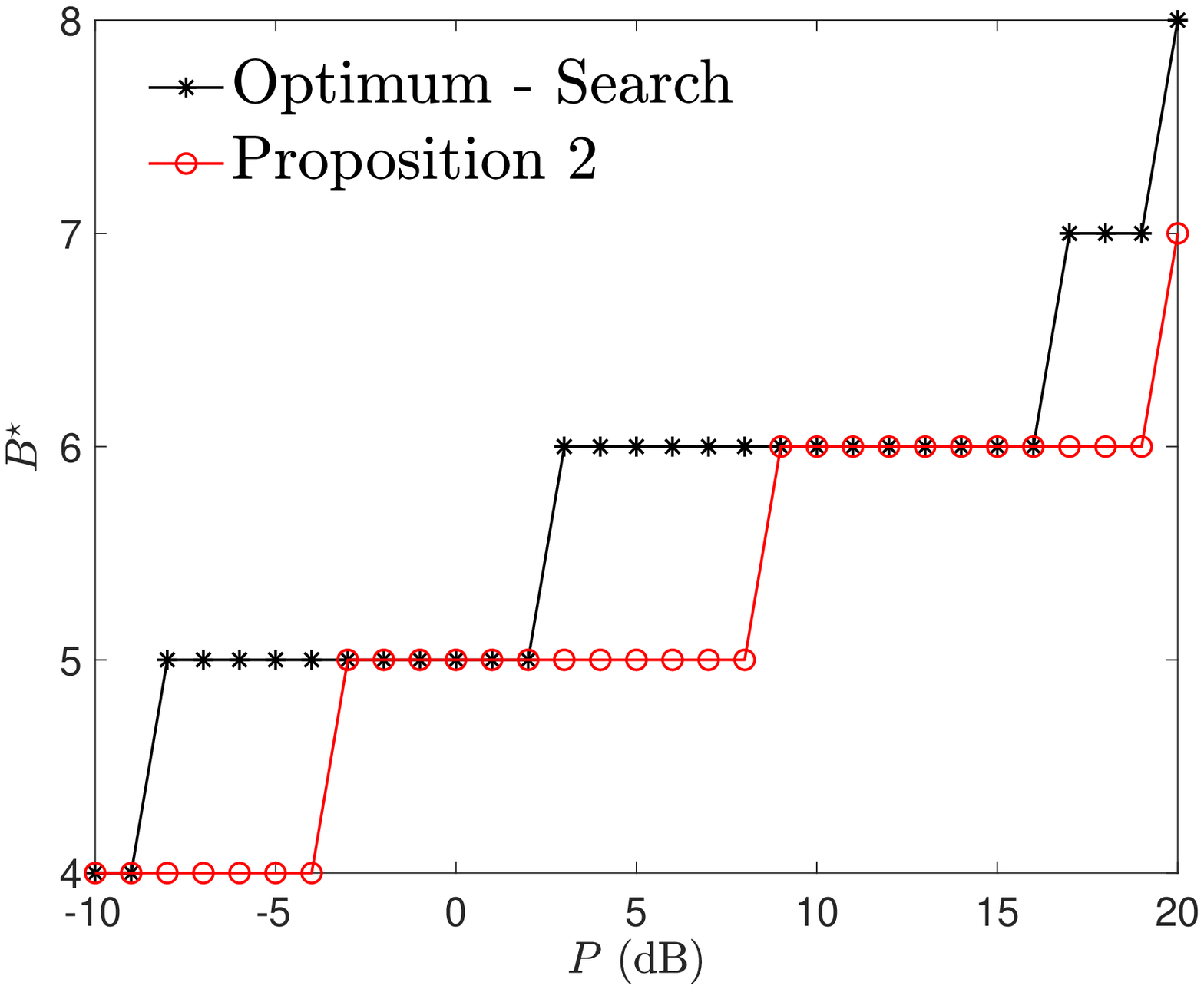}%
\label{fig3.2:a}%
}\hfil
\subfigure[  ]{%
\includegraphics[width=0.5\linewidth]{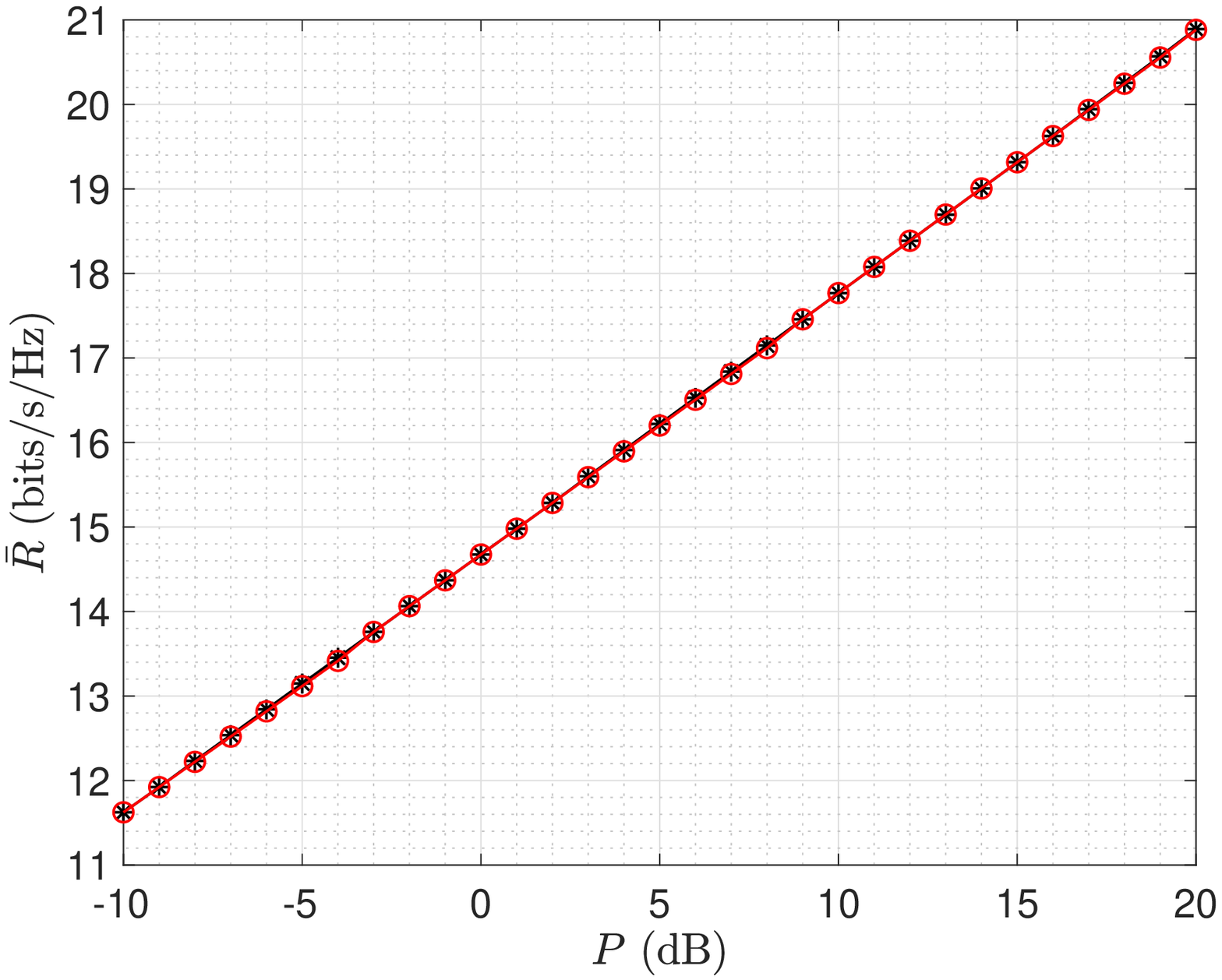}%
\label{fig3.2:b}%
}
\caption{The plots show (a) the optimal group size $B^{\star}$, and (b) the corresponding $\Bar{R}$ as a function of transmit power $P$. The plots show $B^{\star}$ obtained from the optimal brute force search, and using the analytical formula in Proposition \ref{thm1}. The main system parameters are $T_c=900$, $K=360$, and the other parameters are defined in the \emph{System Parameters} of Fig. \ref{fig3.1}.
}
\label{fig3.2}
\end{figure}

\begin{figure}[ht] 
\centering
\subfigure[ ]{%
\includegraphics[width=0.5\linewidth]{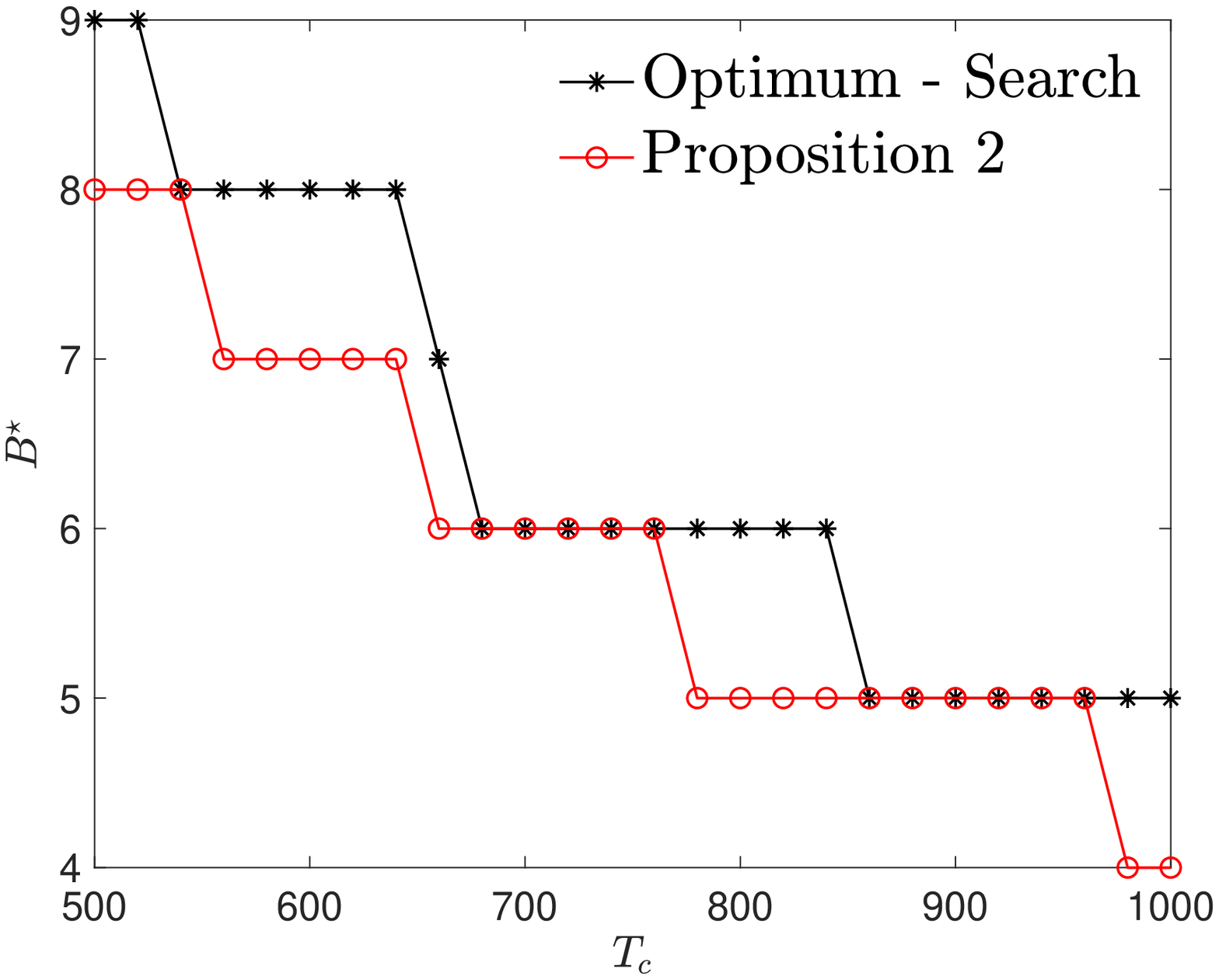}%
\label{fig3.3:a}%
}\hfil
\subfigure[ ]{%
\includegraphics[width=0.5\linewidth]{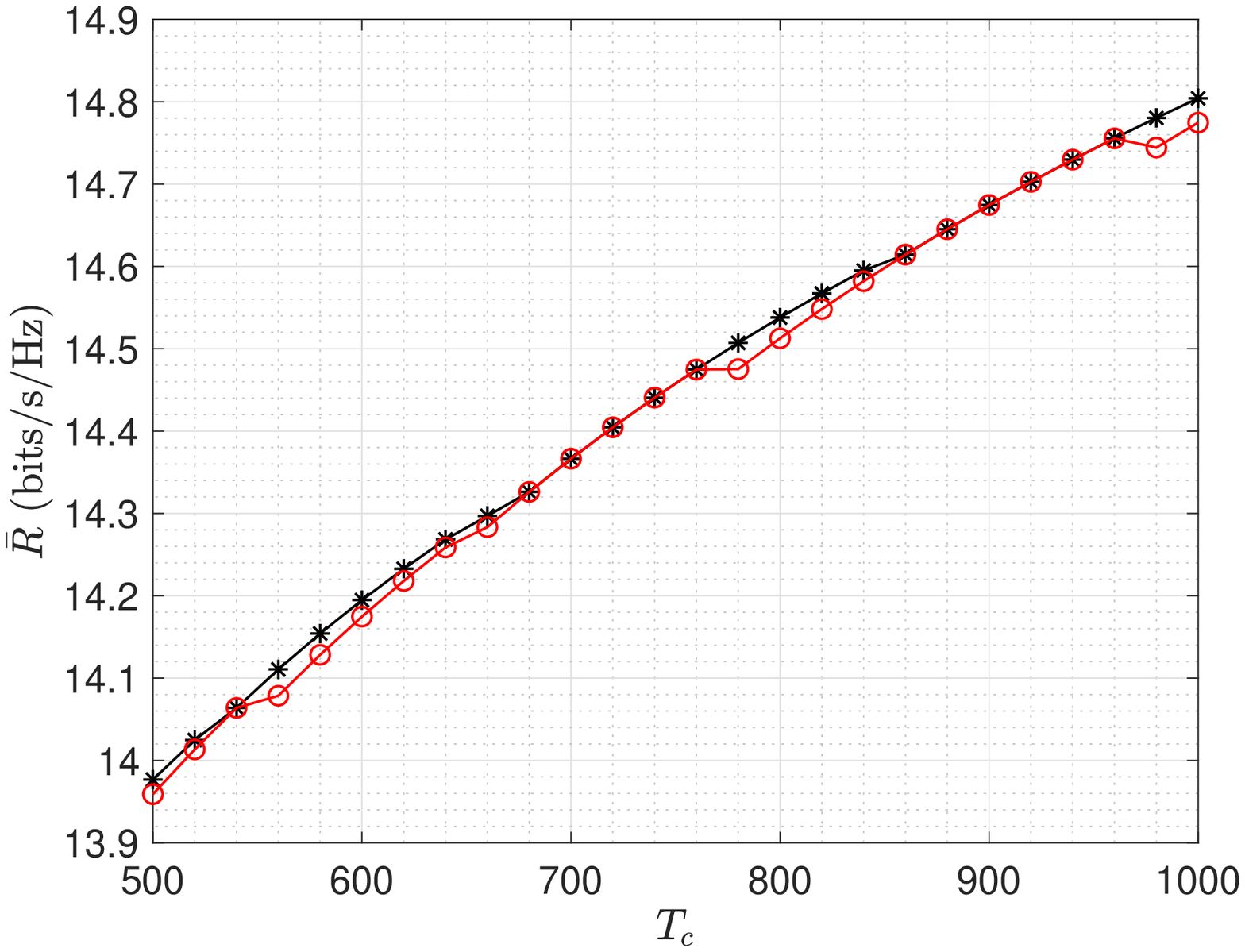}%
\label{fig3.3:b}%
}

\caption{ The plots show (a) the optimal group size $B^{\star}$, and (b) the corresponding $\Bar{R}$ as a function of coherence time $T_c$. Results are shown based on the optimal brute force search and the analytical formula in Proposition \ref{thm1}. The main system parameters are $P=0$ dB, $K=360$, and the other simulation parameters are defined in the \emph{System Parameters} of Fig. \ref{fig3.1}.
}
\label{fig3.3}
\end{figure}

Next, we check the accuracy of Proposition \ref{thm1} and study the effects of varying $P$ and $T_c$ on the optimal group size $B^{\star}$. Fig. \ref{fig3.2:a} plots the optimal group size $B^{\star}$ and Fig. \ref{fig3.2:b} plots the corresponding $\Bar{R}$, both as a function transmit power $P$. The plots show that the $B^{\star}$ obtained from the analytical formula in Proposition \ref{thm1} matches fairly closely with that obtained via a brute-force search, albeit with a small gap since $B^{\star}$ can take only positive integer values.
However, the resulting difference in $\Bar{R} \left(B^{\star} \right)$ is negligible, as evident from Fig. \ref{fig3.2:b}. The plots in Fig. \ref{fig3.2:a} also show that $B^{\star}$ monotonically increases with $P$, which is consistent with the mathematical analysis presented in the previous section. Next we plot the optimal group size $B^{\star}$ and the corresponding $\Bar{R}$ as a function of $T_c$ in Fig. \ref{fig3.3}. As before, although there is some gap between the closed form solution from Proposition \ref{thm1} and the optimum $B^{\star}$, the corresponding $\Bar{R}$ are almost the same. The optimal $B^{\star}$ monotonically decreases with $T_c$, which is consistent with the mathematical analysis in the previous section. 

The results of this paper can guide RIS system design by providing a rough estimate of the number of RIS elements to be grouped together. For a typical outdoor scenario in the sub $6$ GHz 5G band, $T_c\sim 500$ \cite{torres2021lower}, which corresponds to a group size of $B^{\star} \approx 8$ or $K' \approx 45$ subgroups from Fig.\ \ref{fig3.3:a}. For the same $T_c$, analysis of the `On-Off' scheme in \cite{kundu2021large} would suggest that $K\approx 70$ should be switched on. Similarly, the optimal $B^{\star}$ can be obtained for other 5G use case scenarios. For a high mobility scenario, $T_c$ is small, hence a higher group size is required to reduce the pilot overhead. Further, if $P$ is large, the effective channel gain is better which reduces the requirement of passive beamforming gain from the RIS, therefore a higher group size is desired to reduce the pilot overhead.


\section{Conclusion}
We have studied the optimal grouping strategy for RIS-assisted communications. Grouping nearby elements (which share a common phase shift) leads to a smaller pilot overhead as compared to the case when all the elements have distinct independent phase shifts. We have derived a tight upper bound for the average achievable rate, and an analytical expression for the optimal group size which depends only on the statistical channel and system parameters. The RIS-controller can determine the optimal group size which can remain constant over multiple channel coherence blocks since statistical parameters change over a long time scale. Our results reveal that the grouping strategy is most important when the channel coherence time is low (high mobility scenarios) or the transmit power is large. Finally, our results show that grouping has a better performance than the `On-Off' scheme \cite{kundu2021large}, since the grouping strategy reflects more power by utilizing all elements within a group. 

Future extensions of this work could incorporate the effect of correlations between closely packed RIS elements on the optimal group size, and investigate hardware grouping strategies where the nearby elements are physically connected.

\bibliographystyle{IEEEtran}
\bibliography{IEEEabrv,RIS_Grouping}
\end{document}